\documentclass[prl,aps,epsf,amsfonts,floats,twocolumn,amssymb,amsmath,groupedaddress,showpacs,floatfix,nofootinbib]{revtex4}

\usepackage{graphicx}
\usepackage[]{latexsym}
\usepackage{bm}

\newcommand{\be}{\begin{equation}}\newcommand{\ee}{\end{equation}}
\newcommand{\bea}{\begin{eqnarray}}\newcommand{\eea}{\end{eqnarray}}
\newcommand{\brr}{\begin{array}}\newcommand{\err}{\end{array}}
\newcommand{\bit}{\begin{itemize}}\newcommand{\eit}{\end{itemize}}
\newcommand{\ben}{\begin{enumerate}}\newcommand{\een}{\end{enumerate}}

\newcommand{\ba}{\begin{array}}
\newcommand{\ea}{\end{array}}

\def\lab{\label}
\def\lf{\left}

\def\non{\nonumber}\def\ran{\rangle}

\def\ri{\right}

\def\te{\theta}

\def\1{{_{1}}}\def\2{{_{2}}}

\def\noHe0{:\;\!\!\;\!\!:H_e(0):\;\!\!\;\!\!:}
\def\noHm0{:\;\!\!\;\!\!:H_\mu(0):\;\!\!\;\!\!:}

\def\lab{\label}

\def\lf{\left}

\def\non{\nonumber}
\def\ran{\rangle}

\def\ri{\right}

\def\te{\theta}

\def\1{{_{1}}}\def\2{{_{2}}}

\begin{document}

\title{Dark matter and dark energy induced by condensates}

\author{ Antonio Capolupo\footnote{e-mail address: capolupo@sa.infn.it}}

 \affiliation{  Dipartimento di Fisica E.R.Caianiello and INFN gruppo collegato di Salerno,
  Universit\'a di Salerno, Fisciano (SA) - 84084, Italy}

\pacs{ }

\begin{abstract}

It is shown that the vacuum condensate induced by many phenomena behaves as a perfect fluid which, under particular conditions, has zero or negative pressure.
 In particular, the condensates of thermal states, of fields in curved space and of mixed particles have been analyzed. It is shown that the thermal states with the cosmic microwave radiation temperature, the Unruh and the Hawking radiations give negligible contributions to the critical energy density of the universe, while the thermal vacuum of the intercluster medium could contribute to the dark matter, together with the vacuum energy of fields in curved space-time and of mixed neutrinos. Moreover, a component of the dark energy can be represented by the vacuum of axion-like particles mixed with photons and   superpartners of neutrinos.
The formal analogy among the systems characterized by the condensates can open new scenarios in the possibility to detect the dark components of the universe in table top experiments.

\end{abstract}

\maketitle

\section{I. Introduction}

Many independent experimental data  \cite{CMBR}-\cite{SNeIa}, support the hypothesis according to which the today observed universe has an accelerating expansion due to an unknown form of energy, called dark energy, which has a negative pressure. The dark energy  represents approximately $68\%$ of the total matter--energy density of the universe and it is distributed isotropically throughout the universe.
Moreover, astronomical observations relative to the speed of rotating galaxies indicate that an unknown form of matter, not interacting electromagnetically, is need in order to  permits the stability of the galaxies and of  cluster of galaxies. The dark matter
makes up about $27\%$ of the universe.
Different models have been proposed to solve the dark energy \cite{odintsovreport}- \cite{odintsovreport12} and dark matter puzzles \cite{kam}- \cite{kam6}, however, the explanation of the dark components of the universe represents still a very big challenge.

Apparently separate research lines regard  the study of  physical systems  characterized by  vacuum condensates
\cite{Hawking:1974sw}-\cite{Birrell}, such as for example the Hawking or the Casimir effect \cite{Casimir,Casimir1}.
%
%
Such  phenomena
 have  a non-zero vacuum energy which cannot be removed by use of the normal ordering procedure.
This fact, in a supersymmetric context, induces  the spontaneous breaking of supersymmetry \cite{Capolupo:2010ek}-\cite{Mavromatos1}.

In the present  paper it is shown that
these interesting issues are intimately bound together in such way that
the vacuum condensate energy can provide contributions   to the  dark energy
  and to the dark matter.
It is shown that all the condensates have state equations depending on the particular regime one considers (high momentum (UV) and  low momentum (IR) regime). The particular cases of the vacuum energy induced by thermal states, by curved spaces and by the mixing of particles are analyzed in detail.

One shows that the thermal states, the Hawking and the Unruh radiations do not contribute significatively to the energy of the universe.
Only the vacuum energy induced by the Hawking effect of hypothetical primordial black holes and
the thermal vacuum of the intracluster medium (ICM), i.e. the vacuum of hot plasma present at the center of a galaxy cluster, can represent a dark matter component.
The vacuum of fields in curved space has a similar  behavior \cite{maroto}.

A further discussion deserves the particle mixing phenomena, i.e. the  neutrino and quark mixing in the fermion sector, and the axion photon mixing
and the meson mixing in the boson sector.
%
%
For such systems, starting from the results of our previous works  \cite{Capolupo:2006et,Capolupo:2008rz,Capolupo:2007hy3}, and from the ones presented in Refs.\cite{Mavromatos1,Mavromatos2}, obtained in a supersymmetric context,
 one shows that the flavor neutrino vacuum can give a contribution to the dark matter with a value compatible with its estimated upper bound, while,  the  quark condensate, because of  the quark confinement inside the hadrons should
 not interact gravitationally.
Moreover, one shows that the condensate of mixed boson
\cite{Blasone:2001du,Capolupo:2004pt},  as axions and axion like particles (ALPs)  in their interaction with photons, and of superpartners of mixed neutrinos, can  contribute to the dark energy with the state equation of the cosmological constant.
It is expected that mixed particles like kaons, $B^{0}$, $D^{0}$ mesons and $\eta-\eta^{\prime}$ system do not  contribute to the energy on large scale, since they are unstable and  not elementary particles.

We point out that, the common origin of the non trivial vacuum energy contributions  of the systems above described resides in the fact that the physical vacuum of these systems is a condensate of couples of particles and antiparticles which lift the zero point energy to a positive value.
%
The formal analogy (Bogoliubov transformations) among the disparate phenomena generating condensates could allow to simulate  the systems here analyzed by means of phenomena as   the superconductivity, the Casimir effect and the Schwinger effect, which   are reproducible  in table top experiments.

In the computations a particular care has been placed to the renormalization procedure since different schemes may provide different renormalized expressions. The choice of the scheme to adopt is imposed by the need to preserve the symmetries of the system considered.  For example, to preserve the Lorentz invariance of the Minkowski space time,
  the dimensional regularization has been used in ref.\cite{Akhmedov}.
   In this way, one has $p_{vacuum} = - \rho_{vacuum}$, which is the cosmological constant equation. On the contrary, in the case of curved spaces, as the Friedmann-Robertson-Walker background, the Lorentz invariance is no more a symmetry of the  metric and a cut-off regularization can be utilized.
In Ref.\cite{maroto}  a comoving cut-off on the momenta has been proposed.
Notice that the cut-off regularization  represents a valid choice also for the other vacuum condensates since the Lorentz invariance is not a symmetry of such systems. Motivated by such facts, in the following the cut-off regularization will be used for all the phenomena considered, apart from the thermal vacuum for which the regularization is not needed.

The paper is structured as follows,
in Sec.II, the Bogoliubov transformations in QFT are introduced and the  condensate structure of the transformed vacuum is shown. In Sec.III, it is presented the general form of the energy density and pressure of vacuum condensates for boson and fermion fields. In Sec.IV it is  analyzed the contribution given to the energy of the universe by thermal states, with reference to the Hawking and Unruh effects. In Sec.V the fields in curved space are considered, and in Sec.VI the contributions to the dark energy and to the dark matter given by  the particle mixing phenomena are presented. Sec.VII is devoted to the conclusions.

\section{II. Bogoliubov transformation and vacuum condensate}

The  Bogoliubov transformations in the  QFT context \cite{Umezawa:1993yq}
describe disparate phenomena such as the Hawking-Unruh effect \cite{Hawking:1974sw,Unruh:1976db}, the Schwinger effect \cite{Schwinger:1951nm}, the BCS theory of superconductivity \cite{Bardeen:1957mv},  the Thermo Field Dynamics \cite{Takahasi:1974zn}, the particle mixing phenomena \cite{Blasone:2002jv}-\cite{Capolupo:2007hy}, the QFT in curved spacetimes \cite{Birrell} and so on.
%
%
For bosons and fermions they are given by
\bea\label{Bog1}\non
 {\tilde a}_{\mathbf{k}}( \xi, t) &=& U^{B}_{\mathbf{k}} \, a_{\mathbf{k}}(t) - V^{B}_{-\mathbf{k}}  \,a^{\dagger}_{-\mathbf{k}}(t)\,,
\\
 {\tilde a^{\dag}}_{-\mathbf{k}}( \xi,  t) &=& U^{B *}_{-\mathbf{k}} \, a^{\dag}_{-\mathbf{k}}(t) - V^{B *}_{\mathbf{k}}  \,a_{\mathbf{k}}(t)\,,
\eea
and
\bea
\label{Bog2}
 {\tilde \alpha}^r_{\mathbf{k}}(\xi, t) &=& U^{F}_{\mathbf{k}} \, \alpha^r_{\mathbf{k}}(t) + V^{F}_{-\mathbf{k}} \, \alpha^{r\dagger}_{-\mathbf{k}}( t)\,, \non
\\
 {\tilde \alpha}^{r\dagger}_{-\mathbf{k}}(\xi,  t) &=& U^{F *}_{-\mathbf{k}} \, \alpha^{r\dagger}_{-\mathbf{k}}(t) + V^{F *}_{ \mathbf{k}} \, \alpha^{r}_{\mathbf{k}}(t)\,,
\eea
with $a_{\mathbf{k}}(t) = a_{\mathbf{k}} e^{-i\omega_{k}t}$, $\alpha^r_{\mathbf{k}}(t) = \alpha^r_{\mathbf{k}} e^{-i\omega_{k}t}$, annihilators for bosons and fermion fields, respectively, such that $a_{\mathbf{k}}|0\rangle_{B} =  \alpha^r_{\mathbf{k}}|0\rangle_F =0$ and $\omega_{\mathbf{k}}=\sqrt{k^2 + m^2}$.
The  coefficients satisfy the conditions
$
U^{B}_{\mathbf{k}} = U^{B}_{-\mathbf{k}}$, $ V^{B}_{\mathbf{k}} = V^{B}_{-\mathbf{k}}$, $|U^{B}_{\mathbf{k}}|^2 - |V^{B}_{\mathbf{k}}|^2 = 1\,,
$
for bosons, and
$
U^{\psi}_{\mathbf{k}} = U^{F}_{-\mathbf{k}}$, $ V^{F}_{\mathbf{k}} = -V^{F}_{-\mathbf{k}}$, $|U^{F}_{\mathbf{k}}|^2 + |V^{F}_{\mathbf{k}}|^2 = 1\,,
$
for fermions. Thus, they have the general form:
$U^{B}_{\mathbf{k}} = e^{i\gamma_{1\mathbf{k}}}\cosh\eta_{\mathbf{k}}(\zeta)$, $ V^{B}_{\mathbf{k}} =e^{i\gamma_{2\mathbf{k}}}\sinh\eta_{\mathbf{k}}(\zeta)$, $U^{F}_{\mathbf{k}} =e^{i\phi_{1\mathbf{k}}}\cos\xi_{\mathbf{k}}(\zeta)$,
$ V^{F}_{\mathbf{k}} =e^{i\phi_{2\mathbf{k}}}\sin\xi_{\mathbf{k}}(\zeta)$, respectively.
The parameter $\zeta$ controls the physics underlying the transformation. For example, $\zeta$ is related to the temperature $T$ in the  Thermo Field Dynamics case, or to the acceleration of the observer in the Unruh effect case.
The phases $\phi_{i\mathbf{k}}$, $\gamma_{i\mathbf{k}}$, with $i=1,2$,  are irrelevant in our discussions.

The  transformations (\ref{Bog1}) and (\ref{Bog2}) can be written at any time $t$ in terms of the generator $J_{\lambda}(\xi,   t)$ as:
\bea
 {\chi}^r_{\mathbf{k}}( \xi,  t) &=& J_{\lambda}^{-1} (\xi, t)\,\chi^r_{\mathbf{k}}(t) J_{\lambda}(\xi,  t)\,,
\eea
with $\chi= a, \alpha$ and $\lambda = B,F$. Similar relations hold for the creation operators. The generators
have  the property $J_{\lambda}^{-1}(\xi)=J_{\lambda}(-\xi)$.
The vacua $ |0( \xi,  t)\rangle_{\lambda}$ annihilated by the new annihilators are related to the original ones $|0\rangle_{\lambda}$  by the formal relations
$ |0( \xi,  t)\rangle_{\lambda} = J^{-1}_{\lambda}(\xi,  t)|0\rangle_{\lambda} $, with $\lambda = F , B$.

This is a unitary operation if $\mathbf{k}$ assumes a discrete range of values, which happens in quantum mechanics in which there is a finite or countable number of canonical (anti-) commutation relations CCRs. In this case, the Fock spaces built on the two vacua are equivalent and any vector in one space can be expressed in terms of a well defined sum of vectors in the other space.   But in QFT, $\mathbf{k}$ assumes a continuous infinity of values, then one has, for bosons,
\bea\non
|0(\xi,t) \rangle_{B} &=& \exp\lf[ -\delta(\mathbf{0})\int\, d^3 \mathbf{k}\; \log\cosh\xi_{\mathbf{k}} \ri]
 \\ &\times & \exp\lf[ \int\, d^3 \mathbf{k}\; \tanh\xi_{\mathbf{k}} (a_{\mathbf{k}}^{\dagger})^2\ri] |0\rangle_{B} \,,
\eea
which is not a unitary transformation any more \cite{Umezawa}. This shows  that the vacuum $| 0 (\xi,t) \rangle_{B}$ cannot be expressed as a superposition of vectors in the Fock space built over $|0\rangle_{B}$. The same is true for the whole Fock space built over $| {0}(\xi,t  )\rangle_{B}$, i.e. the two Fock spaces are unitarily inequivalent \cite{Umezawa}. Similar discussion holds for fermions.

In general, the existence in QFT of infinitely many representations which are unitarily inequivalent to each other, leads to the problem of the right choice of the Fock space and of the physical vacua associated with the particles which appear in observations.
For the mentioned systems  \cite{Hawking:1974sw}--\cite{Birrell}, ruled by Bogoliubov transformations, the physical vacua  to be used in the computations are the $|0(\xi,t) \rangle_{\lambda}$ ones \cite{Umezawa}.

Notice that, $|0(\xi,t)\rangle_{\lambda}$  is a condensate of couples of particles and antiparticles. Indeed one has
\bea\label{cond1}
 &&_{\lambda}\langle 0(\xi,t)| \chi^{\dagger}_{\mathbf{k}} \chi_{\mathbf{k}} |0(\xi,t)\rangle_{\lambda}   = |V_{\mathbf{k}}^{\lambda}|^2;\,\,\,\,\,\,\, \label{cond2}
\eea
where $\chi =a, \alpha$, $\lambda =B,F$ and Eqs.(\ref{Bog1})--(\ref{Bog2}) have been used.
Such a condensate structure   leads to an energy momentum tensor different from zero for $|0(\xi,t)\rangle_{\lambda} $.

\section{III. Energy-momentum tensor of vacuum condensate}

One considers the free energy momentum tensor densities $T^{\mu\nu}(x)$ for real scalar fields $\phi$ $,
T_{B}^{\mu\nu}(x) = \partial_{\mu}\phi(x) \partial^{\mu}\phi(x) -\frac{1}{2}  g_{\mu\nu} (\partial^{\rho}\phi(x) \partial_{\rho}\phi(x) - m^{2}\phi(x)^{2})\,,
$ and for Majorana fields $\psi$,
$
T_{F}^{\mu\nu}(x) = \frac{i}{2}  \bar{\psi}(x)\gamma_{\mu} \overleftrightarrow{\partial}^{\mu}\psi(x)  \,.
$
One computes the expectation value of   $T^{\lambda}_{\mu \nu}(x)$, ($\lambda =B,F$)  on the transformed vacuum
$| 0 (\xi,  t)\rangle_{\lambda}$,
\bea
\Xi^{\lambda}_{\mu \nu}(x) & \equiv & _{\lambda}\langle 0 ( \xi, t)|: T^{\lambda}_{\mu \nu}(x): | 0 (\xi,  t)\rangle_{\lambda}
\\\non
& = & _{\lambda}\langle 0 ( \xi, t)| T^{\alpha}_{\mu \nu}(x) | 0 (\xi,  t)\rangle_{\lambda}
- _{\lambda}\langle 0  | T^{\lambda}_{\mu \nu}(x) | 0  \rangle_{\lambda}\,.
\eea
The simbol, $:...:$,  denotes the normal ordering with respect to the original vacuum $|0 \rangle_{\lambda}$.
Notice that the off-diagonal components of the expectation value of $T^{\lambda}_{\mu \nu}(x)$ in Eq.(\ref{tensor}) are zero, $\langle 0 ( \xi, t)|: T^{\lambda}_{i,j}(x): | 0 (\xi,  t)\rangle = 0$, for $i \neq j$,
being different from zero only the diagonal components. This implies that the condensates induced by Bogoliubov transformations behave as a perfect fluid and  the  energy density and pressure of boson and fermion  condensates  can be defined as
\bea
\rho^{\lambda} & = &  \langle 0 ( \xi, t)|: T_{0 0}^{\lambda}(x): | 0 (\xi,  t)\rangle\,,
\\
p^{\lambda} & = & \langle 0 ( \xi, t)|: T_{j j}^{\lambda}(x): | 0 (\xi,  t)\rangle\,,
\eea
respectively.

In the boson case, one has
\begin{widetext}

\bea \label{T00Bos}
\rho_{B} & = &  \frac{1}{2}  \langle 0 ( \xi, t)|   : \Big[\pi^{2}( x) + \lf(\vec{\nabla} \phi( x)  \ri)^{2}
+ m^{2} \phi^{2}( x) \Big]: |0( \xi, t) \rangle\,;
\\ \label{TjjBos}
p_{B} & = &    \langle 0 ( \xi, t)| :\Big(\lf[ \partial_{j} \phi( x) \ri]^{2}+ \frac{1}{2}\Big[\pi^{2}( x)
 -   \lf(\vec{\nabla} \phi( x)  \ri)^{2}
 - m^{2}
\phi^{2}( x)  \Big] \Big): |0 ( \xi, t) \rangle\,.
 \eea

\end{widetext}

 In the particular case of the isotropy of the momenta, $k_1 = k_2 =k_3$, one has,
 $ \lf[ \partial_{j} \phi( x) \ri]^{2} = \frac{1}{3} \lf[\vec{\nabla} \phi( x)  \ri]^{2}$, then the pressure can be written as

\begin{widetext}

 \bea
\label{Press-Bos}\non
p_{B} =  \frac{1}{2}   \langle 0 ( \xi, t)| : \Big[\pi^{2}( x)
 -  \frac{1}{3}  \lf(\vec{\nabla} \phi( x)  \ri)^{2}
- m^{2} \phi^{2}( x)  \Big]  : |0 ( \xi, t) \rangle\,.
 \eea

\end{widetext}

In general, the vacuum condensates $|0 ( \xi, t) \rangle_{\lambda}$ are space or time dependent, therefore they violate the Lorentz invariance. This fact implies that  the kinetic and gradient terms of Eq.(\ref{Press-Bos}) can be different from zero. Namely,
$\langle 0 ( \xi, t)|: \pi^{2}( x) : | 0 (\xi,  t)\rangle \neq 0$ and
$\langle 0 ( \xi, t)|:\lf[\vec{\nabla} \phi( x)  \ri]^{2} : | 0 (\xi,  t)\rangle \neq 0$.
Therefore, from  Eqs.(\ref{T00Bos}) and (\ref{Press-Bos}), one can have the following state equations: $w_{B} = p_{B}/\rho_{B}= 1$, if the kinetic term dominates;
$w_B = -1/3$, if the gradient term dominates, and $w_B = -1$, (cosmological constant state equation) for dominating mass term. Moreover, the radiation state equation, $w_B = 1/3$ can be achieved if the kinetic and gradient terms are of the same order, and the mass term is negligible. Such a result is achieved in the high momenta regime, $k \rightarrow \infty$. Finally, the state equation of the dark matter, $w_B = 0$, is obtained for negligible gradient term and for kinetic and mass terms of the same order. This situation happens in the low momenta regime $k \rightarrow  0$.

If all the terms (kinetic, gradient and mass ones) are taken into account, the energy density and pressure are given by
\bea\label{enBos1}
 \rho_{B} & = &  \int \frac{d^{3} {\bf k}}{(2 \pi)^{3}}\, \omega_{k }\, \langle 0 (  \xi, t)| a^{\dagger}_{\bf k}\, a_{\bf k}\, |0( \xi, t) \rangle\,;
\\\non
\\ \label{PreBos}\non
p_{B} & = &   \int \frac{d^{3} {\bf k}}{(2 \pi)^{3}}\,  \Big[\frac{1}{3} \frac{k^2}{\omega_k}  \langle 0 ( \xi, t)|    a^{\dagger}_{\bf k}\, a_{\bf k}\,|0 (  \xi, t) \rangle
\\\non
& - & \lf( \frac{1}{3}  \frac{k^2}{\omega_k} + \frac{1}{2} \frac{m^2}{\omega_k} \ri)
 \langle 0 (  \xi, t)|\Big( a_{\bf k}\, a_{-\bf k} e^{-i \omega_k t}\,
\\
& + &
 a^{\dagger}_{\bf k}\, a^{\dagger}_{ -\bf k} e^{ i \omega_k t}\Big)   |0 ( \xi, t) \rangle \Big]\,,
 \eea
 which, explicitly become
\bea\label{energy-Bos}
\rho_{B}   & = & \frac{1}{2 \pi^{2}}  \int_{0}^{\infty} dk k^{2}\omega_{k } |V_{ k}^{B}|^{2}\,,
\\\non
\\\label{pressure-Bos}\non
p_{B}  & = & \frac{1}{6 \pi^{2}}   \int_{0}^{\infty} dk k^{2}\, \Big[  \frac{k^2}{\omega_k}
|V_{ k}^{B}|^{2}
\\
& - &  \lf( \frac{  k^2}{ \omega_k} + \frac{3 m^2}{2 \omega_k} \ri) |U_{ k}^{B}||V_{ k}^{B}| \cos (\omega_{k} t)  \Big] \,.
 \eea

The state equation is then
\bea\non
w_{B} & = & \frac{1}{3}\frac{\int d^{3}\, \mathbf{k} \frac{k^2}{\omega_k}|V_{k}^{B}|^2 }{\int d^{3} \mathbf{k}\, \omega_k |V_{k}^{B}|^2}
\\
& - &   \frac{1}{3}\frac{\int d^{3} \mathbf{k}\, \lf( \frac{  k^2}{ \omega_k} + \frac{3 m^2}{2 \omega_k} \ri)  U_{k}^{B} V_{k}^{B} \cos (\omega_k t) }{\int d^{3} \mathbf{k}\, \omega_k |V_{k}^{B}|^2}\,.
\eea

In the fermion case, the  energy density and   the pressure   are

\bea \label{T00Ferm}\non
\rho_{F} & = &  \frac{1}{2}   \langle 0 ( \xi, t)|   : \Big[-i \bar{\psi}\, \gamma_{j} \partial^{j}\, \psi + m \bar{\psi} \psi \Big]: |0( \xi, t) \rangle\,;
\\
\\ \label{TjjFerm}
p_{F} & = &     \langle 0 (\xi, t)| :\Big(  \frac{i}{2} \bar{\psi} \,\gamma_{j} \overleftrightarrow{\partial_{j}} \psi  \Big): |0 (\xi, t) \rangle\,.
 \eea
In Eq.(\ref{T00Ferm}), one used the relation $\frac{i}{2} \bar{\psi} \,\gamma_{0} \overleftrightarrow{\partial_{0}} \psi = i \bar{\psi} \,\gamma_{0}  {\partial_{0}} \psi = -i \bar{\psi}\, \gamma_{j} \partial^{j}\, \psi + m \bar{\psi} \psi$. 
By considering the following form of the energy momentum tensor density
\bea\non
:T_{F}^{\mu\nu} : & = & :\lf[\frac{i}{2}  \bar{\psi} \gamma_{\mu} \overleftrightarrow{\partial}_{\nu}\psi
- \eta_{\mu\nu} \lf(\frac{i}{2}  \bar{\psi} \gamma_{j} \overleftrightarrow{\partial}^{j}\psi - m \bar{\psi} \psi \ri) \ri]:
\eea
one can obtain for fermion vacuum condensates the same state equations achieved for boson condensates (see above).
When all the terms are considered (kinematic, gradient and mass ones), from Eqs.(\ref{T00Ferm}) and  (\ref{TjjFerm}), one has
\bea \label{T00Fer1}\non
\rho_{F} & = &   \sum_{r} \int \frac{d^{3} {\bf k}}{(2 \pi)^{3}} \lf(\frac{k^{2}}{\omega_{k}}+ \frac{m^{2}}{\omega_{k}}  \ri) \langle 0 ( \xi, t)|   \alpha^{r \dag}_{k}  \alpha^{r  }_{k}|0(\xi, t) \rangle\,;
\\
\\ \label{TjjFer1}
p_{F} & = &   \frac{1}{3}\sum_{r} \frac{d^{3} {\bf k}}{(2 \pi)^{3}}  \frac{k^{2}}{\omega_{k}}  \langle 0 ( \xi, t)|   \alpha^{r \dag}_{k}  \alpha^{r  }_{k}|0( \xi, t) \rangle\,,
 \eea
where $ \alpha^{r}_{k}$ is the annihilator of fermion field, with $r =1,2$.

 The explicit expressions of the energy density and pressure  are
\bea\label{energy-Ferm}
\rho_{F}  &= & \frac{1}{\pi^{2}} \int_{0}^{\infty} dk k^{2} \omega_{k }  |V_{ k}^{F}|^{2}\,,
 \\\label{pressure-Ferm}
 p_{F}    &= & \frac{1 }{3 \pi^{2}}    \int_{0}^{\infty} dk \frac{k^{4}}{\omega_{k }}|V_{ k}^{F}|^{2}\,.
 \eea
The state equation is then
\bea
w_{F} & = & \frac{1}{3}\frac{\int d^{3}\, \mathbf{k} \frac{k^2}{\omega_k}|V_{k}^{F}|^2 }{\int d^{3} \mathbf{k}\, \omega_k |V_{k}^{F}|^2}\,.
\eea

The origin of   the non-zero   $\rho_{\lambda}$ and $p_{\lambda}$ is due to the fermionic and bosonic
condensates structure of the physical vacuum  which  lift  the vacuum energy and pressure by positive amounts.
Notice also that, being $J^{-1}(\xi,t)  =  J^{\dag}(\xi,t) = J(-\xi,t)$, one can write
\bea\label{tensor}\non
&&_{\lambda}\langle 0 ( \xi, t)|: T^{\lambda}_{\mu \nu}(x): | 0 (\xi,  t)\rangle_{\lambda} =
\\&& = _{\lambda}\langle 0  |J^{-1}_{\lambda}(-\xi,t) : T^{\lambda}_{\mu \nu}(x): J_{\lambda}(-\xi,t)| 0  \rangle_{\lambda} \,.
\eea

Such a property will be used in the following and the notation  $\Theta(-\xi,x) = J^{-1}(-\xi,t) \Theta(x) J(-\xi,t) $ will be adopted to denote the operators  transformed by the generator $J (-\xi,t)$.

Systems such as the thermal states and the particle mixing phenomenon, have  the generators $J's$ which satisfy the condition, $ \vec{\nabla} J(\xi,t) = 0 $. Then, the energy density and pressure for bosons and fermions become,

\begin{widetext}

\bea \label{T00Bos1}
\rho_{B}   & = &  \frac{1}{2} \langle 0| : \Big[\pi^{2}(-\xi,x) + \lf[\vec{\nabla} \phi(-\xi,x)  \ri]^{2}
+ m^{2} \phi^{2}(-\xi,x) \Big]: |0 \rangle\,;
\\ \label{TjjBos1}
p_{B}  & = &   \langle 0| :\Big(\lf[ \partial_{j} \phi(-\xi,x) \ri]^{2}+ \frac{1}{2}\Big[\pi^{2}(-\xi,x)
 -   \lf[\vec{\nabla} \phi(-\xi,x)  \ri]^{2}
 - m^{2}
\phi^{2}(-\xi,x)  \Big] \Big): |0 \rangle\,,
 \eea
\end{widetext}

\bea\label{T00Fer1}\non
\rho_{F} &=& - i   \langle 0| : \Big[\psi^{\dag}(-\xi,x)\gamma_{0} \gamma^{i} \partial_{i} \psi(-\xi,x)
\\
&&+ m
\psi^{\dag}(-\xi,x) \gamma_{0}  \psi(-\xi,x)\Big]: |0 \rangle\,,
\\\non
\\
\label{TjjFer1}
p_{F} &=&  i  \langle 0| :\Big[\psi^{\dag}(-\xi,x)\gamma_{0} \gamma_{j} \partial_{j} \psi(-\xi,x) \Big]: | 0 \rangle \,,
 \eea
respectively.
Eqs.(\ref{T00Bos1})-(\ref{TjjFer1}) will be used to describe the vacuum contributions of  the mixed particles.
Notice that Eqs.(\ref{T00Bos1}), (\ref{TjjBos1}) and (\ref{T00Fer1}), (\ref{TjjFer1}) do not coincide with the more general
Eqs.(\ref{T00Bos}), (\ref{TjjBos}) and (\ref{T00Ferm}), (\ref{TjjFerm}), respectively, since, as a rule, the operator  $\nabla$  and the generators $J's$ do not commute.
%
%

  Eqs.(\ref{T00Bos}), (\ref{TjjBos}) and (\ref{T00Ferm}), (\ref{TjjFerm}), hold for disparate physical phenomena. The explicit form of the Bogoliubov coefficients $U^{\lambda}_{\textbf{k}}$ and $V^{\lambda}_{\textbf{k}}$, $\lambda = B, F$,
specifies the particular system. In the following only few phenomena will be considered in detail. However, the formal analogy among the systems characterized by the condensates, i.e. the Bogoliubov transformations, permits to extend the discussions contained in the next sections also to different phenomena, some of which reproducible in laboratory.

\section{IV. Thermal states, Hawking and Unruh effects}

We consider the formal framework of the Thermo Field Dynamics (TFD) \cite{Takahashi:1974zn}--\cite{Umezawa:1993yq},
which has been successful applied to a number of physical problems at non-zero temperature, in condensed matter physics, in nuclear physics, in particle physics and cosmology.
In the case of systems at non-zero temperature, the physical vacuum is the thermal vacuum state $|0(\xi(\beta))\rangle_{\lambda}$, with $\lambda = B,F$,   $\beta \equiv 1/(k_{B}T)$ and $k_{B}$ the Boltzmann constant \cite{Takahashi:1974zn}--\cite{Umezawa:1993yq}.  The state  $|0(\xi(\beta))\rangle_{\lambda}$ is defines in such a way that the thermal statistical average ${\cal N}_{\chi_{\bf k}}(\xi)$ is given by
  ${\cal N}_{\chi_{\bf k}}(\xi) = _{\lambda}\langle 0(\xi(\beta))| N_{\chi_{\bf k}} |0(\xi(\beta))\rangle_{\lambda}$, with $N_{\chi_{\bf k}} = \chi^{\dag}_{\bf k} \chi_{\bf k} $, ($\chi = a, \alpha$) the number operator \cite{Takahashi:1974zn,Umezawa}. In the boson case,
  $|0(\xi(\beta))\rangle_{B}$ is expressed as
\bea
\lab{(2.12)} |0(\xi(\beta))\rangle_{B} = \prod_{\bf k}
{1\over{\cosh{\xi_k}}} \exp{ \left ( \tanh {\xi_k} ~a_{\bf
k}^{\dagger} {b}_{\bf k}^{\dagger} \right )} |0\rangle_{B} \,,
\eea
where $|0\rangle_{B}$ is  the vacuum annihilated by $a_{\bf k}$ and $b_{\bf k}$.
The auxiliary boson operator $b_{\bf k}$  commutes with $a_{\bf k}$  and is introduced in order to produce the trace operation in computing thermal averages. Similar discussions hold for fermions.

 The thermal vacuum $| 0(\xi(\beta))  \rangle_{\lambda}$  is normalized to one,   and  in the infinite volume
limit one has $
{_{\lambda} \langle 0(\xi(\beta)) | 0 \rangle _{\lambda} \rightarrow 0~~ {\rm
as}~~ V\rightarrow \infty }, ~~~\forall~  \beta\,
$.
Moreover, for  $\beta' \neq \beta$, one has  $ {_{\lambda} \langle 0(\xi(\beta)) | 0(\xi(\beta'))  \rangle_{\lambda}  \rightarrow 0~
{\rm as}~ V\rightarrow \infty}$. Therefore,  $\{  |0(\xi(\beta)) \rangle_{\lambda} \} $ provides a representation  of
the CCR  defined at each  {\it $\beta$} and  unitarily inequivalent $\forall~ \beta'\neq \beta$ to any other representation $\{ |0(\beta') \rangle_{\lambda} \}$ in the infinite volume limit \cite{Takahashi:1974zn}--\cite{Umezawa:1993yq}.

The annhilation operators of $| 0(\xi(\beta))  \rangle_{\lambda}$ (the tilde operators)  are obtained by means of a Bogoliubov transformations similar to Eqs.(\ref{Bog1}) and (\ref{Bog2}), with $a_{- \bf k}$ and ${\tilde a}_{- \mathbf{k}}(\xi,t)$ replaced for bosons by $b_{ \bf k}$ and ${\tilde b}_{ \mathbf{k}}(\xi(\beta))$, respectively, and $\alpha_{- \bf k}$ and ${\tilde \alpha}_{- \bf k}(\xi,t)$ replaced  for fermions by the auxiliary operators $\beta_{  \bf k}$ and  ${\tilde \beta}_{  \bf k}(\xi(\beta))$, respectively. The Bogoliubov coefficients are given by
$U^{T}_{{\bf k}} = \sqrt{\frac{e^{\beta \omega_{\bf k} }}{e^{\beta \omega_{\bf k } }\pm 1}}$ and
$V^{T}_{{\bf k}} = \sqrt{\frac{1}{e^{\beta \omega_{\bf k} } \pm 1}}$, with $-$ for bosons and $+$ for fermions,  $\beta = 1/ k_{B}T$
and $\omega_{\bf k } = \sqrt{k^{2} + m^{2} }$ \cite{Takahashi:1974zn}--\cite{Umezawa:1993yq}.
Such coefficients, used in Eqs.(\ref{energy-Bos}), (\ref{pressure-Bos}) and  (\ref{energy-Ferm}), (\ref{pressure-Ferm}) give the contributions of the thermal vacuum states to the energy  and pressure.
If one considers the  cosmic microwave background temperature, i.e. $T = 2.72 K$, one has that the non-relativistic particles give negligible contribution to the vacuum energy. Only photons and particles with masses of order of $(10^{-3} - 10^{-4})eV$ can contribute significatively to the energy radiation \cite{CapolupoCMB:2016}. Indeed in such cases one obtains energy densities of order of $ 10^{-51}GeV^{4} $ and state equations, $w   =   1/3$ \cite{CapolupoCMB:2016}.

The thermal states can describe also the  Unruh  and of the Hawking effects.
The temperature  is $T = \frac{\hbar a}{2 \pi c k_{b}}$  for Unruh effect, with $a$ acceleration of the observer,  and   $T=\frac{\hbar c^{3}}{8 \pi G M k_{b}}$, for Hawking effect,
with $M$ black hole mass and $G$ gravitational constant.
Both of the  phenomena do not contribute to the energy of the universe since their temperature is very low. For example, a black hole of one solar mass has a temperature of only $60 nK$ and the thermal vacuum energy of any particle is negligible. Only primordial black holes with very small mass could have temperatures higher than the one of the cosmic microwave background and then give a contribution to the energy of the universe.

A non-trivial contribute  can be given by the thermal vacuum of the intracluster medium.
Such a hot plasma filling the center of galaxy clusters has temperatures of order of $(10 \div 100) \times 10^{6}K $.
For example,  the thermal vacuum of free electrons with temperature of $80 \times 10^{6}K $ has an energy of $10^{-47} GeV^4$ and a state equation $w = 0.01$. Such values are in agreement with the ones on the dark matter.

\section{V. Fields in curved background}

Another example of condensed vacuum system is represented by fields in curved spaces. In these cases, the energy momentum tensor for spin $0$ and $1/2$ are given by \cite{Birrell}
\bea\non
T_{\mu \nu}^{s=0}(x) & = & (1 - 2 \xi) \phi_{;\mu} \phi_{;\nu} + (2 \xi - \frac{1}{2}) g_{\mu \nu}g^{\rho\sigma}\phi_{;\rho} \phi_{;\sigma}
\\\non
& - & 2  \xi \phi_{;\mu \nu } \phi + \frac{2}{n} \xi g_{\mu \nu} \phi  \Box \phi
\\ \non
& - & \xi \lf[R_{\mu\nu} - \frac{1}{2}R g_{\mu\nu} + \frac{2(n - 1)}{n}\xi R g_{\mu\nu}  \ri]\phi^{2}
\\
 & + & 2 \lf[ \frac{1}{4} - \lf(1 - \frac{1}{n} \ri)\xi \ri] m^{2} g_{\mu \nu} \phi^{2}\,,
\eea
\bea
T_{\mu \nu}^{s=\frac{1}{2}}(x) = \frac{i}{2}  \lf[ \bar{\psi} \gamma_{(\mu} \nabla_{\nu)}\psi - \lf(\nabla_{(\mu} \bar{\psi} \ri)\gamma_{\nu)}\psi \ri]\,,
\eea
respectively, and the energy density and pressure depend on the particular metric considered.
Let us consider the spatially flat Friedmann Robertson-Walker metric
\bea
d s^{2} = d t^{2} - a^{2}(t) d {\bf x}^{2} = a^{2}(\eta) (d \eta^{2}- d {\bf x}^{2})\,,
\eea
where $a$ is the scale factor, $t$ is the comoving time, $\eta$ is the conformal time, $\eta(t) = \int_{t_{0}}^{t} \frac{d t }{a(t)}$, with $t_0$ arbitrary constant.
The boson field $\phi(\mathbf{x},\eta)$ can be expressed as
\bea
\phi(\mathbf{x},\eta) = \int d^{3} \mathbf{k} \lf[a_{\mathbf{k}} \phi_{k}(\eta)+  a^{\dag}_{-\mathbf{k}}\phi^{*}_{-k}(\eta)  \ri] e^{i \mathbf{k}\mathbf{x}}\,,
\eea
where the mode functions $\phi_{k}(\eta)$ have analytical expression only in particular cases.
%
However, in any case, the energy density and pressure, after the introduction of a cut-off on the momenta, can be written as \cite{Parker1,Parker2}
\bea\non
\rho_{curv} &=& \frac{2 \pi}{a^2}\int_{0}^{K} dk k^{2} \lf(|\phi_{k}^{\prime}|^{2} + k^{2} |\phi_{k}|^{2} + m^{2} |\phi_{k}|^{2} \ri),
\\
\\\non
p_{curv} &=& \frac{2 \pi}{a^2}\int_{0}^{K} dk k^{2} \lf(|\phi_{k}^{\prime}|^{2} - \frac{k^{2}}{3} |\phi_{k}|^{2} - m^{2} |\phi_{k}|^{2} \ri).
\\
\eea
In Ref.\cite{maroto} it has been shown that at late time the cutoff on the momenta can be assumed much smaller than the comoving mass of the field, $K \ll m a$. Moreover, assuming that $m \gg H$,   in an arbitrary Robertson-Walker metric for infrared regime, one obtains \cite{maroto}
\bea\non
\rho_{curv} &=& \frac{1}{8 \pi^2} \int_{0}^{K} dk k^{2} \lf(\frac{2 m}{a^{3}} + \frac{9 H^{2}}{4 m a^{3}} + \frac{k^2}{m a^{5}} \ri),
\\
\\\non
p_{curv} &=& \frac{1}{8 \pi^2}\int_{0}^{K} dk k^{2} \lf( \frac{9 H^{2}}{4 m a^{3}}  - \frac{k^2}{3 m a^{5}} \ri).
\\
\eea
Therefore, state equation is  $w_{curv} \simeq 0$, i.e. the energy of the vacuum of a scalar field in curved space behaves as a dark matter component in the infrared regime. The  energy density is \cite{maroto}
 \bea\label{curv}
 \rho_{curv} = \frac{m K^{3}}{12 \pi^{2} a^{3} }.
 \eea
The value of $\rho_{curv}$ depends on the values of the mass field $m$, on the scale factor $a$ and on the cutoff on the momenta $K$. Thus,   numerical  values compatible with the ones of dark matter  can be found only for the values of the parameters such that,
$\frac{m K^{3}}{ a^{3} } \sim 10^{-45}GeV^{4}$.

We expect  a similar result for fermion fields in curved spaces.
Further study on such topics represents a work in progress.

\section{VI. Particle mixing}

 The  field mixing phenomenon is represented by the mixing of neutrinos and quarks in fermion sector and by the axion-photon mixing and the mixing of kaons, $B^0$, $D^0$, and $\eta-\eta^\prime$ systems, in boson sector. For two fields, it is expressed, both for fermions and bosons, as
  \bea\label{mixing}\non
 \varphi_{1}(\theta,x) &=&     \varphi_{1}(x) \cos(\theta) + \varphi_{2}(x) \sin(\theta)\,,
\\
 \varphi_{2}(\theta,x) &=&  - \varphi_{1}(x) \sin(\theta) + \varphi_{2}(x) \cos(\theta)\,,
 \eea
 where, $\theta$ is the mixing angle, $ \varphi_{i}(\theta,x)$ are the mixed fields and $ \varphi_{i}(x)$ are the free fields, with $i =1,2$.
In the case of neutrino mixing, the mixed fields are the flavor neutrino fields, $\varphi_{1}(\theta,x) \equiv \nu_{e}$, $\varphi_{2}(\theta,x) \equiv\nu_{\mu}$   and the free fields are the neutrinos with definite masses $m_1$ and $m_2$,  $\varphi_{1}(x)\equiv \nu_1$ and $\varphi_{2}(x)\equiv \nu_2$. The two flavor neutrino mixing case has been considered. In the boson sector, for axion--photon mixing in the presence of a magnetic field  one has, $\varphi_{1}(\theta,x) \equiv \gamma_{\|}(z)$, $\varphi_{2}(\theta,x) \equiv a (z)$, (with $\gamma_{\|}(z)$ photon polarization field parallel to the purely transverse magnetic field $\mathbf{B} = \mathbf{B}_{T}$  and $a$ axion field).
The free fields are   $\varphi_{1}(x)\equiv \gamma^{\prime}_{\|}(z)$ and $\varphi_{2}(x)\equiv a^{\prime} (z) $, with
$\gamma^{\prime}_{\|}(z) = \gamma^{\prime}_{\|}(0) e^{- i \omega_{\gamma} z}$, $a^{\prime} (z) = a^{\prime} (0) e^{- i \omega_{a} z}$
and
\[
\omega_{\gamma} = \omega + \Delta_{-}\,, \quad
\omega_{a} = \omega+ \Delta_{+}\,,
\]

\[
\Delta_{\pm} = - \frac{\omega_{P}^{2}+ m_{a}^{2}}{4 \omega} \pm \frac{1}{4 \omega} \sqrt{(\omega_{P}^{2}- m_{a}^{2})^{2} + (2 g \omega B_{T})^2}\,.
\]
The mixing angle is
$\theta_{a} = \frac{1}{2} \arctan \displaystyle{\left(\frac{2 g \omega B_{T}}{m_{a}^{2}-\omega^2_P} \right)}$, where $g  \in [10^{-16}-10^{-10}]GeV^{-1} $ is the axion-photon coupling, $m_{a}\in [ 10^{-6} - 10^{-2}]eV$ is the axion mass and $\omega_P$ is the plasma frequency.
Moreover, for instable mesons,
 the mixed fields are $K^{0}$ and $\bar{K}^{0}$,   $B^{0}$ and $\bar{B}^{0}$, or $D^{0}$ and $\bar{D}^{0}$. The corresponding free fields are $K_{L}$ and $K_{S}$, $B_{L}$ and $B_{H}$, and $D_{L}$ and $D_{H}$, respectively and
$\cos(\theta) =  \frac{1}{2 p} \sqrt{1 - z} $, $\sin(\theta) =  \frac{1}{2 p} \sqrt{1 + z} $,
with $
\frac{q}{p}\,=\, \sqrt{\frac{\mathcal{H}_{21}}{\mathcal{H}_{12}}}\,,
$ (${\mathcal{H}_{ij}}$ are the elements of the
effective Hamiltonian ${\mathcal{H} }$ of mixed meson systems) and $z$ parameter describing the $CPT$ violation.

The mixing transformations (\ref{mixing}) can be written as $\varphi_{i}(\theta,x) \equiv J^{-1}(\theta, t) \varphi_{i}(x) J (\theta, t)$, where $  i  =
1,2$, and $J (\theta, t)$ is the transformation generator \cite{Blasone:2002jv,Blasone:2001du}.
Analogously, the mixed
annihilation operators are  $\chi^{r}_{{\bf k},i}(\theta, t)
\equiv J^{-1}(\theta, t)\;\chi^{r}_{{\bf k},i}(t)\;J (\theta, t)$, with
$\chi^{r}_{{\bf k},i} = a_{{\bf k},i}, \alpha^{r}_{{\bf k},i}$, for bosons and fermion, respectively and $i =1,2$.
 They annihilate the mixed vacuum
$|0(\theta, t)\ran \,\equiv\,J^{-1}(\theta, t)\;|0\ran_{1,2} $, where
$|0\ran_{1,2}$ is the vacuum annihilated by $\chi^{r}_{{\bf k},i} $.

The  physical vacuum where particle oscillations appears  is $|0(\theta, t)\ran $. It is a (coherent) condensate
of $\chi_{{\bf k},i}$ particles
(antiparticles) \cite{Blasone:2002jv,Blasone:2001du}:
\bea \label{con}   \langle 0(\theta, t)| \chi_{{\bf k},i}^{r \dag} \chi^r_{{\bf
k},i} |0(\theta, t)\rangle  = \sin^{2}\te ~ |\Upsilon^{\lambda}_{{\bf
k}}|^{2},
\eea
where $\lambda = B,F$, $i=1,2$ and the reference frame ${\bf k}=(0,0,|{\bf k}|)$ has
been adopted for convenience. $\Upsilon^{\lambda}_{{\bf
k}}$ is the Bogoliubov
coefficient entering the mixing transformation. For boson and fermion one has \cite{Blasone:2002jv,Blasone:2001du}
\bea\label{Bogoliubov}
| \Upsilon^{B}_{{\bf
k}}|  &=&   \frac{1}{2} \lf( \sqrt{\frac{\Omega_{k,1}}{\Omega_{k,2}}} -
\sqrt{\frac{\Omega_{k,2}}{\Omega_{k,1}}} \ri)\,,
\\
|\Upsilon^{F}_{{\bf
k}}|  &=&  \frac{ (\Omega_{k,1}+m_{1}) - (\Omega_{k,2}+m_{2})}{2
\sqrt{\Omega_{k,1}\Omega_{k,2}(\Omega_{k,1}+m_{1})(\Omega_{k,2}+m_{2})}}\, |{\bf k}| \,,
\eea
respectively,
with $| \Sigma^B_{{\bf k}}|^{2}  - | \Upsilon^B_{{\bf k}}|^{2}  = 1 $ and $| \Sigma^F_{{\bf k}}|^{2}  + | \Upsilon^F_{{\bf k}}|^{2}  = 1 $,
($\Sigma^\lambda_{{\bf k}}$ are the other coefficients entering in the transformations), $\Omega_{k,i}$ energies of the free fields, for example,
$\Omega_{k,i} = \omega_{\gamma}, \omega_{a}$ for axions-photon mixing, $\Omega_{k,i} = \omega_{k,i}$ for neutrinos, $i=1,2$.
.

\subsection{Boson mixing}

 For mixed bosons,
 Eqs.(\ref{T00Bos1}) and (\ref{TjjBos1}) become

\begin{widetext}
 \bea \label{T00BosMix}\non
\rho^{B}_{mix} & = &   \langle 0 ( \theta, t)|: T^{B-mix}_{0 0}(x): | 0 (\theta,  t)\rangle
\\
& = & \frac{1}{2} \langle 0| : \sum_i \Big[\pi^{2}_{i}(-\theta,x) + \lf[\vec{\nabla} \phi_{i}(-\theta,x)  \ri]^{2}
+ m^{2}_{i} \phi^{2}_{i}(-\theta,x) \Big]: |0 \rangle\,;
\\\non
\\ \label{TjjBosMix}\non
p^{B}_{mix} & = &   \langle 0 ( \theta, t)|: T^{B-mix}_{j j}(x): | 0 (\theta, x)\rangle
\\
& = &  \langle 0| :\sum_i  \Big(\lf[ \partial_{j} \phi_{i}(-\theta,x) \ri]^{2}+ \frac{1}{2}\Big[\pi_{i}^{2}(-\theta,x)
 -
  \lf[\vec{\nabla} \phi_{i}(-\theta,x)  \ri]^{2}
 - m_{i}^{2}
\phi_{i}^{2}(-\theta,x)  \Big] \Big): |0 \rangle\,,
 \eea
 \end{widetext}
respectively, where
 \bea\non\label{mixing-rel}
 \pi_{1}(-\theta,x) &=&   \pi_{1}(x) \cos(\theta) - \pi_{2}(x) \sin(\theta)\,,
\\
 \pi_{2}(-\theta,x) &=&   \pi_{1}(x) \sin(\theta) + \pi_{2}(x) \cos(\theta)\,.
 \eea
From Eqs.(\ref{mixing-rel}), one can immediately see that,  $\sum_i \pi^{2}_{i}(-\theta,x) = \sum_i \pi^{2}_{i}(x)  $,
  $\sum_i [\vec{\nabla} \phi_{i}(-\theta,x)]^{2} = \sum_i [\vec{\nabla} \phi_{i}(x)]^{2}  $, and $\sum_i [\partial_{j} \phi_{i}(-\theta,x)]^{2}= \sum_i [\partial_{j} \phi_{i}(x)]^{2}$, i.e. such operators are invariant under the action of the generator $J(-\theta,t)$.
  This fact implies that the kinetic and gradient terms of the mixed vacuum are equal to zero
  \bea\non
  && \langle 0| : \sum_i  \pi^{2}_{i}(-\theta,x):|0 \rangle\, =  \langle 0| : \sum_i   \lf[\vec{\nabla} \phi_{i}(-\theta,x)  \ri]^{2}
 : |0 \rangle\,
 \\
 && = \langle 0| :\sum_i  \lf[ \partial_{j} \phi_{i}(-\theta,x) \ri]^{2}:|0 \rangle\, = 0\,.
  \eea
  Then, Eqs.(\ref{T00BosMix}) and (\ref{TjjBosMix}) become
   \bea \label{T00BosMixF}
\rho^{B}_{mix} & = &    \langle 0| :\sum_{i}
  m^{2}_{i}
\phi_{i}^{2}(-\theta,x) : |0 \rangle\,,
\\ \label{TjjBosMixF}
p^{B}_{mix} & = & -    \langle 0| :\sum_{i}
  m^{2}_{i}
\phi_{i}^{2}(-\theta,x) : |0 \rangle\,,
 \eea
    and the state equation is $w^{B}_{mix} = - 1$, (which is the state equation of the cosmological constant), independently on the choice of the cut-of on the momenta.
 Similar result has been obtained in supersymmetric context in Refs. \cite{Mavromatos1,Mavromatos2}.
 Here one analyzes the possible phenomenological implications of such result.
 Denoting with  $\Delta m^{2}= |m_{2}^{2}-m_{1}^{2}|$, the energy density of the boson mixed vacuum  is explicitly given by
 \bea\label{integral}
 \rho^{B}_{mix} = \frac{\Delta m^{2}   \sin ^{2}\theta}{8 \pi^{2}} \int_{0}^{K} dk k^{2} \lf(\frac{1}{\omega_{k,1}}- \frac{1}{\omega_{k,2}}\ri),
 \eea
where $K$ is the cut-off on the momenta.
Eq.(\ref{integral}) will be now solved for
 the mixing of axion-like particles and for the flavor mixing of supersymmetric partners of neutrinos.

- {\it Contribution of axion like particles} - In  the case of the mixing between the photon and the axion-like particles, denoting with $m_1 = m_\gamma = 0 $, the photon mass and with $m_{2} = m_{a}  $, the axion mass, one has
\bea\label{ener-axion}\non
\rho^{axion}_{mix}  &=& \frac{  m_{a}^{2} \sin ^{2}\theta_{a}}{16 \pi^{2} } \Big[K \lf(K -\sqrt{K^{2}+ m_{a}^{2}} \ri)
\\
&+& m_{a}^{2} \log \lf( \frac{K+\sqrt{K^{2}+ m_{a}^{2}}}{m_{a}} \ri)
\Big]\,.
\eea
In astrophysical contexts, as in the case of  active galactic nuclei, quasars, supernova and magnetars,
the magnetic field strength  vary between, $B \in [10^{6} - 10^{17}] G$. Also the plasma frequencies and the photon energies $\omega$ vary considerably. Therefore the mixing angles  $\theta_{a}$ depends by the particular system one consider.

Notice, however that for axion mass $m_{a}$ of order of $2 \times 10^{-2}eV$ and   $\sin^{2}_{a}\theta \sim 10^{-2}$ (which could be obtained for different astrophysical objects), using the Planck scale cut-off, $K\sim 10^{19} GeV$, one has $\rho^{axion}_{mix} = 2.3 \times 10^{-47}GeV^{4}$,
 which is compatible with the estimated upper bound on the dark energy.

 Smaller values of $m_{a}$ or of $\sin^{2}\theta \sim 10^{-2}$, lead to smaller values of  $\rho^{axion}_{mix}$.

- {\it Contribution of neutrino superpartners } -
In the case  superpartners of the neutrinos, the  integral (\ref{integral}) leads to
\bea\label{ener-Bos}\non
\rho^{B}_{mix}  &=& \frac{\Delta m^{2} \sin ^{2}\theta}{16 \pi^{2} } \Big[K \lf(\sqrt{K^{2}+ m_{1}^{2}}-\sqrt{K^{2}+ m_{2}^{2}} \ri)
\\\non
&+& m_{2}^{2} \log \lf( \frac{K+\sqrt{K^{2}+ m_{2}^{2}}}{m_{2}} \ri)
\\
&-& m_{1}^{2} \log \lf( \frac{K+\sqrt{K^{2}+ m_{1}^{2}}}{m_{1}} \ri)\Big]\,.
\eea
One considers then, masses similar to the ones of the neutrinos,  $m_1 = 10^{-3} eV$ and $m_2 = 9 \times 10^{-3} eV$, in order that $\Delta m^{2} = 8 \times 10^{-5} eV^{2}$. Moreover, one assumes
$\sin^{2} \theta = 0.3$.  One obtains, $\rho^{B}_{mix}  = 7 \times 10^{-47}GeV^{4}$ for a cut-off on the momenta  $K =10  eV$, and $\rho^{B}_{mix}  = 6.9 \times 10^{-46}GeV^{4}$  for a cut-off of order of the Planck scale, $10^{19} GeV$.
Smaller values of the mixing angle lead to values which are compatible with the estimated value of the dark energy also in the case in which the cut-off is $K = 10^{19} GeV$, indeed $\rho^{B}_{mix}$ depends linearly by  $\sin^{2}\theta$.


\subsection{Fermion mixing}

In the case of fermion mixing, Eqs. (\ref{T00Fer1}), (\ref{TjjFer1})  become
\begin{widetext}
\bea\label{T00FerMix}
\rho_{mix}^{F} & = & -   \langle 0| : \sum_i \Big[\psi_{i} ^{\dag}(-\theta,x)\gamma_{0} \gamma^{j} \partial_{j} \psi_{i}(-\theta,x)
+ m
\psi_{i}^{\dag}(-\theta,x) \gamma_{0}  \psi_{i}(-\theta,x)\Big]: |0 \rangle\,;
\\\label{TjjFerMix}
p_{mix}^{F}  & = &  i    \langle 0| :\sum_i \Big[\psi_{i}^{\dag}(-\theta,x)\gamma_{0} \gamma_{j} \partial_{j} \psi_{i}(-\theta,x) \Big]: | 0 \rangle \,,
 \eea
 \end{widetext}
 where $\psi_{i}(-\theta,x)$ are the flavor neutrino fields or the quark fields.
Being
\bea\non
\sum_i \bar{\psi}_{i}(-\theta,x)\gamma^{j} \partial_{j} \psi_{i}(-\theta,x) &=& \sum_i \bar{\psi}_{i}(x) \gamma^{j} \partial_{j} \psi_{i}(x)
 \\\non
 \sum_i \Big[\psi_{i}^{\dag}(-\theta,x)\gamma_{0} \gamma_{j} \partial_{j} \psi_{i}(-\theta,x)\Big]&=& \sum_i \Big[\psi_{i}^{\dag}(x) \gamma_{0} \gamma_{j} \partial_{j} \psi_{i}(x) \Big] ,
 \eea
then
\bea\non
 && \langle 0| : \sum_i \bar{\psi}_{i}(-\theta,x)\gamma^{j} \partial_{j} \psi_{i}(-\theta,x): | 0 \rangle \ =
 \\
 && \langle 0| : \sum_i \Big[\psi_{i}^{\dag}(-\theta,x)\gamma_{0} \gamma_{j} \partial_{j} \psi_{i}(-\theta,x)\Big]: | 0 \rangle \ = 0\,.
\eea

Thus  the   energy density and pressures of  Eqs.(\ref{T00FerMix}) and (\ref{TjjFerMix}), become
\bea\label{T00FerMixF}
\rho_{mix}^{F}  & = & -   \langle 0| : \sum_i \Big[m_i
\psi_{i}^{\dag}(-\theta,x) \gamma_{0}  \psi_{i}(-\theta,x)\Big]: |0 \rangle,
\\\label{TjjFerMixF}
p_{mix}^{F}  & = &  0 \,,
 \eea
respectively.
The pressure of the fermion mixed vacuum $| 0 (\theta,  t)\rangle $ is equal to zero independently on the regularization adopted. Then the state equation  in this case is $w^{F}_{mix} =0$, which is the one of the dark matter.
Similar result has been obtained in supersymmetric context in Refs. \cite{Mavromatos1,Mavromatos2}.
Now one shows the consequences of such a result.
Solving Eq.(\ref{T00FerMixF}), the energy density of the fermion mixed vacuum is
\bea\label{ener-Fer}\non
\rho^{F}_{mix} = \frac{\Delta m \sin ^{2}\theta}{2 \pi^{2}} \int_{0}^{K} dk k^{2}
\lf( \frac{m_2 }{ \omega_{k,2}} - \frac{m_1}{ \omega_{k,1}}\ri)\Big]\,,
\eea
and explicitly, one has

\bea\label{ener-Fer}\non
\rho^{F}_{mix} &=& \frac{\Delta m  \sin ^{2}\theta}{2 \pi^{2}} \Big[K \lf(m_{2} \sqrt{K^{2}+ m_{2}^{2}}- m_{1}\sqrt{K^{2}+ m_{1}^{2}} \ri)
\\\non
&-& m_{2}^{3} \log \lf( \frac{K+\sqrt{K^{2}+ m_{2}^{2}}}{m_{2}} \ri)
\\
&+& m_{1}^{3} \log \lf( \frac{K+\sqrt{K^{2}+ m_{1}^{2}}}{m_{1}} \ri)\Big]\,.
\eea

For masses of order of $10^{-3}eV$, such that $\Delta m^{2}$ is of order of $8 \times 10^{-5}eV^2$ and a cut-off on the momenta  $K = m_1 + m_2$, one obtains  $\rho^{F}_{mix} = 4 \times 10^{-47 } GeV^4$, which is in agreement with the estimated upper bound of the dark matter.
A possible mechanism which imposes a very low cut-off  for neutrinos is given in Ref.\cite{Mavromatos3}.
Values of $K$ of order of the Plank scale leads to  $\rho^{F}_{mix} \sim  \times 10^{-46 } GeV^4$.

We point out that we have considered an empty universe without matter and gravitational interaction. This fact produces the homogeneity of the condensates.
The presence of matter interacting  with the neutrino flavor vacuum and with the  vacuum condensates above presented could generate clustered matter  and therefore the irregularities which are observed in the universe.

By considering the quark masses, the value of $\rho^{F}_{mix}$ one obtains is much higher than the ones above obtained.
 However,  the quark confinement inside the hadrons should inhibit the gravitational interaction of the quark vacuum condensate. Therefore such condensate should  play none role in the formation of structures on large scale and it should not affect the dark matter component.

Moreover, the condensates induced by the mixed bosons, which behaves as a cosmological constant, can evolve remaining homogeneous also in the presence of matter, since their pressure is negative.

\section{VII. Conclusions}

It has been shown that the vacuum condensates induced by different phenomena can contribute to the dark sector of the universe. The vacuum states of these systems are indeed condensates of couples of particles and antiparticles which generate non-zero vacuum energies and which, under particular conditions, behave as dark matter or dark energy.

The  contributions given by the thermal states, by the fields in curved space and by the particle mixing phenomenon have been analyzed.
It has been shown that the thermal states and the condensates due to Unruh and Hawking effects do not contribute considerably to the vacuum energy.
Non-trivial contributes to the energy are given by the thermal vacuum of the intercluster medium, by the vacuum of   fields in curved space-time  and by the flavor vacuum of neutrinos. Such vacuum condensates have a negligible pressure, therefore their state equations are similar to the dark matter one. Moreover the values of the energies are compatible with the one estimated for the dark matter.

On the other hand,   the  mixing between photons and axion-like particles can reproduce the behavior and the estimated value of the dark energy component.

The formal analogy existing among completely different phenomena, characterized by vacuum condensates, suggests to investigate the properties
of the condensates of the systems here studied,  in phenomena like the superconductivity, the Casimir effect and the Schwinger effect, which can be analyzed in table top experiments. Therefore, the results presented in  this paper can   open a completely new way  in the research of dark matter and in the study of the dark energy.

\section{Conflict of Interests}

The authors declare that there is no conflict of interests
regarding the publication of this paper.

\section{Acknowledgements}
Partial financial support from MIUR is acknowledged.

\end{document}